\documentclass[11pt]{article}
\usepackage[margin=.68in]{geometry}
\usepackage{amsmath,amssymb,amsthm}
\usepackage[hidelinks]{hyperref}
\usepackage{url}
\newtheorem{remark}{Remark}
\newcommand{\E}{\mathbb E}

\title{A Note on Scaling in Randomly Rotated Quantization and Its Connection to the CDEF $+1$ Pythagorean Relation}

\date{}
\author{
Uri Erez\\
School of Electrical Engineering\\
Tel Aviv University\\
Tel Aviv, Israel\\
uri@eng.tau.ac.il
}

\begin{document}
\maketitle

\begin{abstract}
Quantization schemes based on randomized rotations have recently received
renewed attention, including the roles of MMSE and unbiased  reconstruction scalings.  In this note, we point out the connection to classical results in statistical signal processing and
communication theory.  Specifically, the two reconstruction scales used in the EDEN line of work admit a natural interpretation as finite-dimensional, realization-dependent counterparts of the Wiener and unbiased coefficients in the classical CDEF formulation.
At finite blocklength, the CDEF +1 relation holds pointwise for each rotation realization as an exact geometric (Pythagorean) identity, 
but does not hold after averaging the distortions over the rotation.
The classical SNR relation
$\sf{SNR}_{\rm MMSE}=\sf{SNR}_{\rm MMSE,U}+1$ is recovered as
$d\to\infty$: once the overall scale is handled separately, the empirical coordinate statistics of a randomly rotated
vector approach their i.i.d. Gaussian counterparts, and the
rotation-dependent quantities concentrate.
Importantly, EDEN goes beyond this classical correspondence: for every finite \(d\), its Haar-rotation formulation guarantees exact conditional unbiasedness, a stronger property than the second-order notion of unbiasedness in CDEF.
We further comment on two distinct roles random rotations play in quantization: one is approximate Gaussianization of the coordinates; the other is decorrelation of reconstruction errors across quantization branches.


\end{abstract}
\section{Introduction}
Quantization schemes based on randomized rotations have recently received
renewed attention, primarily due to their central role in LLM quantization;
representative examples include
\cite{chee2023quip,ashkboos2024quarot,zandieh2026turboquant}.
Some of the underlying ideas, and in particular the
random-rotate--quantize--rotate-back (RQRB) pipeline, can be traced back to
classical areas in information theory and statistical signal processing; see,
e.g., the discussions in
\cite{zandieh2026turboquant,ordentlich2026high}.
The purpose of this note is to shed some further light on these connections.
Specifically, we:

\begin{enumerate}
    \item Trace two distinct roles of the RQRB pipeline in the quantization
    literature: approximate Gaussianization on the one hand, and
    decorrelation of quantization errors on the other, with the latter
    corresponding to its interpretation as a dithering mechanism \cite{hadad2016dithered_msc,hadad2016dithered}.

    \item Relate the two reconstruction scalings employed in the EDEN line of
    work \cite{vargaftik2022eden,ben2026note} to the classical theory of
    Cioffi, Dudevoir, Eyuboglu, and Forney (CDEF)
    \cite{cioffi1995mmse}. Namely, the EDEN scalings are precisely
    finite-dimensional, realization-dependent counterparts of the classical
    Wiener and unbiased CDEF scalings. This correspondence is exact for every
    rotation realization; as the dimension grows, concentration connects
    these realization-dependent quantities to the classical scalar-Gaussian
    setting and its familiar $+1$ SNR relation.
\end{enumerate}


\section{The Classical Biased/Unbiased Relation}
\label{sec:II}
We begin with the classical scalar setting.  Let $X$ and $Y$ be correlated
zero-mean random variables, and denote
$P_X=\E[X^2]$, $P_Y=\E[Y^2]$, and $r_{XY}=\E[XY]$.

As we recall next, Cioffi, Dudevoir, Eyuboglu, and Forney (CDEF) distinguish two natural
linear scalings of $Y$, corresponding to biased and unbiased linear MMSE
estimation~\cite{cioffi1995mmse}; see also Forney's later Hilbert-space
exposition~\cite{forney2004shannon}.

\subsection{Wiener scaling}

The Wiener, or linear-MMSE, coefficient minimizes $\E[(X-\alpha Y)^2]$ and is given by
\begin{equation}
    \alpha_{\rm W}=\alpha_{\rm W}(r_{XY},P_Y)
    \triangleq
    \frac{r_{XY}}{P_Y}.
    \label{eq:wiener-scale}
\end{equation}
By the orthogonality principle, the optimal estimation error satisfies
\begin{equation}
    \E[E Y]=0,
    \label{eq:wiener-orthogonality}
\end{equation}
where $\hat X=\alpha_{\rm W}Y$ is the linear MMSE estimator and
$E=X-\hat X$.\footnote{In the sequel MMSE will be used to denote linear MMSE.}

Thus, Wiener scaling makes the estimation error orthogonal to the observation
and yields the backward-channel relation
\begin{equation}
    X=\alpha_{\rm W}Y+E,
    \label{eq:backward-channel}
\end{equation}
with $E \perp Y$.

\subsection{Unbiased (unit signal gain) scaling}

As developed in CDEF \cite{cioffi1995mmse}, the unbiased coefficient is\footnote{More precisely, CDEF begin with the forward-channel relation \eqref{eq:classical-forward-channel} while noting that it can be enforced to hold via scaling.}
\begin{equation}
    \alpha_{\rm U}=\alpha_{\rm U}(r_{XY},P_X)
    \triangleq
    \frac{P_X}{r_{XY}}.
    \label{eq:unit-gain-scale}
\end{equation}
The unbiased estimator is then
\begin{equation}
    Z\triangleq\alpha_{\rm U}Y=X+N,
    \label{eq:classical-forward-channel}
\end{equation}
where
\begin{equation}
    \E[XN]
    =
    \alpha_{\rm U}\E[XY]-\E[X^2]
    =0.
    \label{eq:source-orthogonality}
\end{equation}
Thus $\alpha_{\rm U}$ makes the effective noise orthogonal to the source ($N\perp X$). The relation \eqref{eq:classical-forward-channel} is called the forward-channel realization.
\begin{remark}
    Here “unbiased” is used in the 
    the weaker, second-order CDEF sense of unit signal gain.
    The condition 
    $X\perp N$ does not by itself imply the (conditional) unbiasedness relation $\E [Z|X]=X$, 
    equivalently $\E[N|X]=0$. See discussion of EDEN in the sequel.
\end{remark}

Writing $P_N=\E[N^2]$, the SNR of the unbiased, forward-channel representation is naturally defined
as
\begin{equation}
    {\sf SNR}_{\rm MMSE,U}
    =
    \frac{P_X}{P_N}.
    \label{eq:classical-unbiased-snr}
\end{equation}
If one starts with the forward channel relation $Z=X+N$, the Wiener reconstruction is obtained by the
additional scaling
\[
    \hat X=\alpha_{{\rm W}|{\rm U}}Z,
\]
with reconstruction error
\begin{equation}
    E_\alpha
    =
    X-\alpha Z
    =
    (1-\alpha)X-\alpha N.
\end{equation}
Since $X\perp N$,
\begin{equation}
    \E[E_\alpha^2]
    =
    (\alpha-1)^2P_X+\alpha^2P_N.
\end{equation}
The minimizing coefficient relative to the unit-gain representation is
\begin{equation}
    \alpha_{{\rm W}|{\rm U}}
    =
    \frac{P_X}{P_X+P_N}.
    \label{eq:classical-beta}
\end{equation}
Thus, we have $\alpha_{{\rm W}|{\rm U}}=\frac{\alpha_{\rm W}}{\alpha_{\rm U}}$, 
or equivalently, 
   $\alpha_{\rm W}=\alpha_{{\rm W}|{\rm U}}\alpha_{\rm U}$.
The resulting MMSE is
\begin{equation}
    {\sf MMSE}
    =
    \frac{P_XP_N}{P_X+P_N}.
\end{equation}
Defining\footnote{Indeed, as should be clear from the exposition to follow, the terminology of signal-to-noise ratio to describe this quantity, though ubiquitous, should be interpreted with caution.} 
\begin{equation}
    \sf{SNR}_{\rm MMSE}
    =\frac{P_X}{ {\sf MMSE}},
\end{equation}
gives the biased/unbiased SNR relation of Lemma~2 of CDEF,
\begin{equation}
    \sf{SNR}_{\rm MMSE}
    =\sf{SNR}_{\rm MMSE,U}+1.
    \label{eq:classical-plus-one}
\end{equation}

\subsection{The Pythagorean interpretation}

The identity also has the geometric interpretation emphasized by
CDEF~\cite{cioffi1995mmse}.  The forward representation
\[
    Z=X+N,
\]
with $X\perp N$,
forms one right triangle.  The Wiener estimate
\[
    \hat X=\alpha_{{\rm W}|{\rm U}}Z
\]
is the orthogonal projection of $X$ onto the one-dimensional subspace
spanned by $Z$, so that
\[
    X=\hat X+E,
\]
with $ 
    E\perp Z$ (and also  $ 
    E\perp \hat X$).
This is the second right triangle in CDEF's Fig.~9; Forney develops the same
projection geometry in his Hilbert-space treatment of MMSE
estimation~\cite{forney2004shannon}.  The $+1$ relation is therefore Pythagoras
applied to the forward- and backward-channel decompositions. 
We identify in the sequel these rules within EDEN \cite{vargaftik2022eden}.
Namely, replacing
random variables by fixed vectors and the mean-square inner product by the
Euclidean inner product gives the same two scaling rules for every
realization. 

\subsection{Why the forward-channel SNR is natural for averaging}

The forward-channel realization \eqref{eq:classical-forward-channel}
is the natural viewpoint when several quantized reconstructions are to be linearly
combined as elaborated on in \cite{hadad2016dithered_msc,hadad2016dithered,ostergaard2022incremental}.\footnote{CDEF studies a channel coding scenario rather than the dual quantization scenario discussed in this note. The exposition here follows that of \cite{ostergaard2022incremental} where the notation ${\sf SDR_{opt}}$ and ${\sf uSDR_{opt}}$ was used in place of ${\sf SNR_{MMSE}}$ and ${\sf SNR_{MMSE,U}}$, respectively.}  
To that end, let us now relate the framework to the problem of averaging independently quantized elements of a vector $(X_1,\ldots,X_m)$, where the elements are correlated. To further simplify the exposition assume the extreme case where all entries are identical; see Section~II.B in \cite{hadad2016dithered}. Namely, we now identify \begin{equation}
   Y_i=Q_i(X_i)=Q_i(X) 
\end{equation} as the output of the $i$-th quantizer $Q_i(\cdot)$. Ideally, we would like to conclude that
\[
Z_i=X+{N}_i,\qquad i=1,\ldots,m,
\]
where $Z_i=\alpha_{{\rm U},i}Y_i$. Assume that the $N_i$ are zero mean,
have equal power $\E[N_i^2]=P_N$, and satisfy $\E[XN_i]=0$ and $\E[N_iN_j]=0$ for $i\ne j$. 
Under these assumptions, their average
is again a forward channel with unit signal gain,
\begin{equation}
\overline{{Z}}=\frac1m\sum_{i=1}^m{Z}_i
 ={X}+\overline{{N}},   
\label{eq:non=coherent}
\end{equation}
 where
 \[
\overline{{N}}=\frac1m \sum_{j=1}^m{N}_j,
\]
and
\[
\E \left[ \overline{{N}}^2\right]=\frac1 {m^2} \sum_{j=1}^m \E[N_i^2]=P_N/m.
\]
We therefore obtain
\begin{equation}
 {\sf SNR}_{\rm MMSE,U,ave}=m {\sf SNR}_{\rm MMSE,U}.
 \label{eq:snr-averaging}
\end{equation}
Thus, the forward-channel representation is natural for averaging, since
the estimation errors behave as additive, zero-mean, mutually uncorrelated
noises. We next discuss several mechanisms for realizing this model.



\subsection{Comments on the Roles of the Random Rotate-Quantize-Rotate Back Pipeline}

Classical subtractive dithering realizes this forward-channel model for a
uniform quantizer without overload, using uniform dither over one quantization
interval: the noise is independent of the input, and independent dithers make
the noises in different descriptions independent \cite{gray1998quantization}.

In \cite{hadad2016dithered_msc,hadad2016dithered}, the random rotate--quantize--rotate-back (RQRB)  pipeline was proposed as an alternative form of dither for scalar quantization, 
with the goal of asymptotically eliminating second-order error correlations
in distributed quantization of correlated sources. Both Haar rotations
and randomized Hadamard transforms are studied, with the latter providing
a computationally efficient implementation.
The sources in that work are already modeled as temporally i.i.d. Gaussian (e.g., following transform coding), so RQRB is not used for Gaussianization. Rather, different random rotations are applied across quantization branches, making the resulting quantization errors asymptotically uncorrelated (via random rotations) and uncorrelated from the source (via CDEF scaling);  hence enabling noncoherent combining.

Shortly afterward, Suresh et al. \cite{suresh2017distributed} proposed using the RQRB pipeline in the closely related context of distributed mean estimation, with the random rotation playing a different role.
A common randomized Hadamard transform is used across clients (branches in the terminology of \cite{hadad2016dithered}) to transform a deterministically modeled source vector so as to control the dynamic range of the elements to be fed into stochastic scalar quantizers. With independent stochastic-rounding randomness across clients, the errors
are zero mean and pairwise uncorrelated, ensuring noncoherent combining. 


In turn, EDEN \cite{vargaftik2022eden} employs RQRB with an independent Haar rotation for each client and deterministic scalar quantization. The rotation makes the relevant empirical coordinate statistics of an arbitrary input vector asymptotically Gaussian, permitting scalar quantization optimized for a Gaussian source, while a realization-dependent scaling ensures  unbiasedness in the strong sense (see Remark 1). Independence across client rotations is
not needed for Gaussianization or single-client unbiasedness.  It is
used at the averaging stage: together with conditional unbiasedness,
it makes the reconstruction errors conditionally independent and
eliminates their cross correlations.  Thus, in EDEN, the rotations
serve both as a Gaussianizing transform and, in the second-order sense
of~\cite{hadad2016dithered}, as dither enabling noncoherent combining.

\section{EDEN Scaling as a Deterministic Realization-Dependent Counterpart of CDEF}

Let  $\mathbf X \in\mathbb R^d$ be a given non-zero vector and consider a realization of a
Haar rotation $U$. The EDEN scheme forms the rotate--quantize--rotate-back observation $\mathbf Y$ before
reconstruction scaling as
\[
 \mathbf Y=\eta_{\mathbf X}^{-1}U^{\mathsf T}Q(\mathbf V),
\]
where
\[
 \eta_{\mathbf X}=\frac{\sqrt d}{\|\mathbf X\|},
\]
and
\[
 \mathbf V=\eta_{\mathbf X}U\mathbf X.
\]
For this deterministic pair, define the correlation
\[
 r_{\mathbf{XY}}^{(d)}=\frac1d\langle\mathbf X,\mathbf Y\rangle,
\]
and normalized energies $P_{\mathbf X}^{(d)}=\frac1d\|\mathbf X\|^2$,
 $
 P_{\mathbf Y}^{(d)}=\frac1d\|\mathbf Y\|^2$.

The two CDEF scalings of Section~\ref{sec:II} then become
\begin{equation}
 \alpha_{\rm W}
 =\alpha_{\rm W}\!\left(r_{\mathbf{XY}}^{(d)},P_{\mathbf Y}^{(d)}\right)
 =\frac{\langle\mathbf X,\mathbf Y\rangle}{\|\mathbf Y\|^2},
 \label{eq:scales}
\end{equation}
and
\begin{equation}
 \alpha_{\rm U}
 =\alpha_{\rm U}\!\left(r_{\mathbf{XY}}^{(d)},P_{\mathbf X}^{(d)}\right)
 =\frac{\|\mathbf X\|^2}{\langle\mathbf X,\mathbf Y\rangle}.
 \label{eq:scales2}
\end{equation}
Indeed,
\[
 r_{\mathbf{XY}}^{(d)}
 =P_{\mathbf X}^{(d)}
   \left(\frac1d\sum_{i=1}^dV_iQ(V_i)\right),
\]
and hence, when EDEN is expressed as a scaling of $\mathbf Y$,
\begin{equation}
 \alpha_{\rm U}
 =\frac{d}{\sum_{i=1}^dV_iQ(V_i)}.
 \label{eq:eden-scale}
\end{equation}
Equivalently, the coefficient multiplying $U^{\mathsf T}Q(\mathbf V)$ is
$\eta_{\mathbf X}^{-1}\alpha_{\rm U}
=\|\mathbf X\|^2/\langle U\mathbf X,Q(\mathbf V)\rangle$, which is
the scale used in the original EDEN formulation \cite{vargaftik2022eden}.

EDEN assumes that the reconstruction scale is represented without
quantization error.
It immediately follows that 
\begin{equation}
 \mathbf Z=\alpha_{\rm U}\mathbf Y=\mathbf X+\mathbf N,
 \label{eq:pointwise}
\end{equation}
with $\mathbf X \perp \mathbf N$,
for every rotation realization, while
$\widehat{\mathbf X}=\alpha_{\rm W}\mathbf Y$ is the
corresponding Euclidean projection.  These are precisely the deterministic
counterparts of the two CDEF scalings. 

Treating $U$ as random, Theorem 2.1 in \cite{vargaftik2022eden} additionally proves the stronger
finite-dimensional (unbiasedness) statement:
\begin{equation}
 \E_U[\mathbf Z]=\mathbf X.
 \label{eq:unbiasedness3}
\end{equation}
Equivalently, if $\mathbf X$ is viewed as random, and independent of $U$,
then $\E[\mathbf Z\mid\mathbf X]=\mathbf X$.
EDEN’s realization-dependent scaling enforces exact source–error orthogonality for every realization; rotational symmetry additionally yields exact conditional unbiasedness.

\subsection{SNR Interpretation and Relation to the Stochastic Scalar Setting}
Define
\[
 A(U)=\frac{\|\mathbf X\|^2\|\mathbf Y\|^2}
 {\langle\mathbf X,\mathbf Y\rangle^2}.
\]
The normalized errors of the unit-gain and Wiener reconstructions are given, respectively, by
\begin{equation}
    D_{\rm U}(U)
=\frac{\|\alpha_{\rm U}\mathbf Y-\mathbf X\|^2}{\|\mathbf X\|^2}
=A(U)-1,
\end{equation}
and
\begin{equation}
    D_{\rm W}(U)
=\frac{\|\alpha_{\rm W}\mathbf Y-\mathbf X\|^2}{\|\mathbf X\|^2}
=1-\frac1{A(U)}.
\end{equation}
Note that, for every realization, they indeed satisfy the CDEF Pythagorean relation:
\[
 \frac{1}{D_{\rm W}(U)}
 =
 \frac{1}{D_{\rm U}(U)}+1.
\]

Define the averaged distortions over $U$  by
$D_{\rm U}^{(d)}=\E_U[D_{\rm U}(U)]$ and
$D_{\rm W}^{(d)}=\E_U[D_{\rm W}(U)]$.
Define further the corresponding two SNRs by
\[
{\sf SNR}_{\rm MMSE,U}^{(d)}=\frac{1}{D_{\rm U}^{(d)}}
\]
and
\[
{\sf SNR}_{\rm MMSE}^{(d)}=\frac{1}{D_{\rm W}^{(d)}}
\]
Concavity of $t/(1+t)$ implies that by Jensen:
\begin{equation}
 {\sf SNR}_{\rm MMSE}^{(d)}
 \geq {\sf SNR}_{\rm MMSE,U}^{(d)}+1.
 \label{eq:jensen-plus-one}
\end{equation}

The large-$d$ interpretation is as follows.  For a Haar rotation,
\[
 \mathbf V\mathrel{\stackrel{\rm d}=}
 \frac{\sqrt d\,\mathbf G}{\|\mathbf G\|},
\]
where
\[
 \mathbf G\sim\mathcal N(\mathbf 0,I_d).
\]
As $d$ grows, $\|\mathbf G\|/\sqrt d$ concentrates around one, and the
empirical quantities entering $A(U)$ approach their scalar Gaussian
counterparts. Indeed,
\[
 A(U)
 =
 \frac{\frac1d\sum_{i=1}^d Q(V_i)^2}
 {\left(\frac1d\sum_{i=1}^d V_iQ(V_i)\right)^2}.
\]
Hence, if
$a=\E[GQ(G)]$, $b=\E[Q(G)^2]$,
 and $G\sim\mathcal N(0,1)$,
then $A(U)$ approaches $b/a^2$.

Consequently, the
realization-dependent EDEN scalings approach the corresponding CDEF
scalings for the scalar Gaussian model, and the Jensen gap vanishes.
We therefore recover the CDEF relation asymptotically:
\[
 {\sf SNR}_{\rm MMSE}^{(d)}
 -{\sf SNR}_{\rm MMSE,U}^{(d)} \longrightarrow 1,
\]
as $d\to\infty$. 

\section*{Acknowledgment}
The author thanks Or Ordentlich for drawing attention to the EDEN line of work and, in particular, to its reconstruction scaling.
\bibliographystyle{IEEEtran}
%

\end{document}